\documentclass[nofootinbib,aps,preprint,longbibliography]{revtex4-1}

\usepackage{gensymb}
\usepackage{color}
\usepackage[normalem]{ulem}
\usepackage{graphicx}
\usepackage{dcolumn}
\usepackage{bm}
\usepackage{amsmath}

\newcommand{\be}{\begin{equation}}
\newcommand{\ee}{\end{equation}}
\newcommand{\bfig}{\begin{figure}}
\newcommand{\efig}{\end{figure}}

\newcommand{\LRSG}{LaRu$_3$(Si$_{1-x}$Ge$_x$)$_2$}
\newcommand{\LRS}{LaRu$_3$Si$_2$}
\newcommand{\LRG}{LaRu$_3$Ge$_2$}

\begin{document}
\title{Chemical enhancement of superconductivity in \LRS{} with \\mode-selective coupling between kagome phonons and flat bands
}
\author{Ryo Misawa$^{1}$}
\author{Markus Kriener$^{2}$}
\author{Rinsuke Yamada$^{1}$}
\author{Ryota Nakano$^{1}$}
\author{Milena Jovanovic$^{2,3}$}
\author{Leslie M. Schoop$^{2}$}
\author{Max Hirschberger$^{1,4}$}\email{hirschberger@ap.t.u-tokyo.ac.jp}

\affiliation{$^{1}$Department of Applied Physics and Quantum-Phase Electronics Center (QPEC), The University of Tokyo, Bunkyo, Tokyo 113-8656, Japan}
\affiliation{$^{2}$Department of Chemistry, Princeton University, Princeton 08540, New Jersey, USA}
\affiliation{$^{3}$Department of Chemistry, North Carolina State University, Raleigh, 27695, North Carolina, USA}
\affiliation{$^{4}$RIKEN Center for Emergent Matter Science (CEMS), Wako, Saitama 351-0198, Japan}
\date{\today}

\begin{abstract}
In kagome metals, flat electronic bands induced by frustrated hopping are a platform for strong electron correlations. In particular, a selective coupling of flat band states to certain kagome phonon modes is proposed as a universal origin of superconductivity in this material class. Here, we investigate the superconductivity in the kagome system \LRSG{} by chemical pressure tuning while preserving the Ru-$4d$ states that constitute the kagome flat bands. We observe a sizable enhancement in the density of states up to $x = 0.07$, as determined by the specific heat, with a concomitant increase in the superconducting transition temperature $T_\mathrm{c}$. Ge-dopants induce a uniaxial lattice expansion along the $c$-axis. Our first-principles calculations suggest that this mitigates the detrimental effect of hybridization between kagome layers and reduces the dispersion of the Ru-$d_{x^2-y^2}$ flat band. The calculated chemical potential moves closer to the maximum in the energy-dependent density of states. Our result is consistent with a theoretical prediction of tunable flat band superconductivity in \LRS{} by mode-selective coupling between specific kagome phonons and the Ru-$d_{x^2-y^2}$ orbitals.\\
\end{abstract}

\maketitle

\begin{figure}[h!]
  \begin{center}
		\includegraphics[clip, trim=0cm 0cm 0cm 0cm, width=0.45\linewidth]{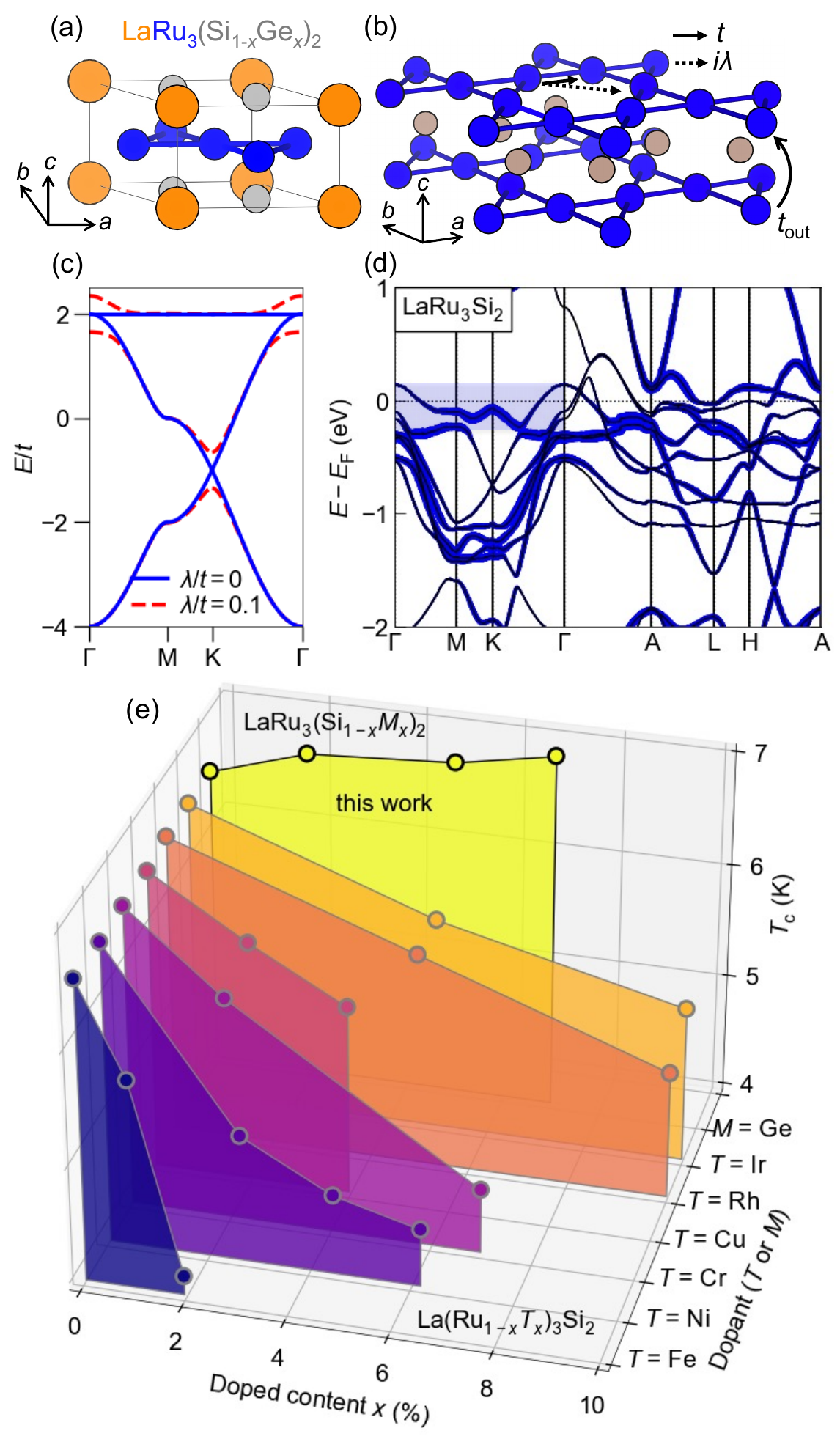}
    \caption[]{(color online). Superconducting kagome metal \LRSG{} (LRSG) with a flat band at the Fermi level.
    (a) Hexagonal structure of superconducting kagome metal LRSG. (b) The kagome motif of Ru atoms with Si and Ge between Ru layers. $t$ and $\lambda$ are the nearest-neighbor hopping and spin-orbit coupling, respectively. Uniaxial lattice expansion along $c$-axis suppresses out-of-plane hopping $t_\mathrm{out}$. 
    (c) Band dispersion of a single-orbital model on the kagome lattice, with Dirac cone and flat band (blue). 
    The introduction of spin-orbit coupling opens a gap between the Dirac cone and the flat band (red).
    (d) Band structure of \LRS{}. Blue shading highlights a flat band in the $k_x-k_y$ plane. Thick blue color: density of states for the Ru-$4d_{x^2-y^2}$ orbital.
    (e) Superconducting transition temperature $T_\mathrm{c}$ for alloyed \LRS{}~\cite{footnote1}.
    This work demonstrates the enhancement of $T_\mathrm{c}$ by chemical pressure tuning, in contrast to previous attempts of band-filling tuning.
    } 
    \label{Fig1}
  \end{center}
\end{figure}

The kagome lattice is a two-dimensional network of corner-sharing triangles with distinct properties in terms of its electronic band topology~\cite{Yin2022-ps,Wang2023-jl,Checkelsky2024-ik}, electron-electron correlations, and phonon spectra~\cite{Deng2025}.
This Letter discusses phonon-driven superconductivity in the intermetallic \LRS{}, where the kagome lattice is formed by Ru atoms as depicted in Fig.~\ref{Fig1}(a). Considering first a single orbital model on an isolated kagome lattice in Fig.~\ref{Fig1}(b),
one obtains three bands forming a flat band and Dirac cone 
[Fig.~\ref{Fig1}(c)]. In particular, the dispersionless upper band originates from frustrated hopping between the three sites of the kagome lattice~\cite{Bergman2008-zr}.
Kane-Mele type spin-orbit coupling~\cite{Kane2005-ja} further opens a gap and imparts a Chern number onto the flat band~\cite{Guo2009-ql,Bolens2019-cq,Kang2020-sh}. 

The single-orbital model is not sufficient to describe the topological indices and band dispersion of real-world kagome metals~\cite{Jiang2025}, but such materials retain key features of this simplified electronic structure if the coupling between consecutive kagome layers is weak
~\cite{Lin2018-zv,Kang2019-hl,Kang2020-sh,Li2021-up,Regnault2022-kz,Ye2024-nr}. 
As an example, Fig.~\ref{Fig1}(d) presents the electronic band structure of \LRS{}, which hosts an hourglass-shaped dispersion originating from Ru-$4d$ orbitals along the $\Gamma$–$K$–$M$ path.
\LRS{} is a rare kagome superconductor with a flat band near the Fermi level, yet without any magnetic instability of said flat band~\cite{Mielke2021-ha,Deng2025}. In \textit{ab-initio} theory, the flat band states couple selectively to certain acoustic kagome phonons~\cite{Deng2025},  discussed further at the end of this Letter, and realize the highest observed transition temperature $T_\mathrm{c} = 6.7\,$K among the family of $RT_3M_2$ ($R$ = rare-earth, $T$ = transition metal and $M$ = main group element)~\cite{Barz1980-ep,Ku1980-nl,Chaudhary2023-do,Gui2022-mq,Rauchschwalbe1984-wk,Mielke2021-ha,Gong2022-jj} and among kagome metals as a material class~\cite{Deng2025}.
At higher temperature, other (optical) phonon modes of \LRS{} soften and cause charge-density-wave (CDW) orders~\cite{Deng2025}, possibly accompanied by breaking of time-reversal symmetry~\cite{Plokhikh2024-eg,Mielke2024-as}. The phenomenology is reminiscent of the kagome superconductors $AM_3$Sb$_5$ ($A =$K, Cs, and Rb; $M=$V, Cr, ...)~\cite{Wilson2024-ca,Ortiz2019-pr}. In that material family, the most closely related compound is CsCr$_3$Sb$_5$, which hosts flat bands at $E_\mathrm{F}$ and a charge density wave (CDW) with the same modulation vector as \LRS{}~\cite{Liu2024-cx}; however, all these phenomena occur at much lower temperature in CsCr$_3$Sb$_5$.

Real kagome metals are three-dimensional, and their flat bands often exhibit strong dispersion along the $k_z$ direction of reciprocal space~\cite{Kang2020-sh,Huang2022-ka,Ye2024-nr}. This behavior is also observed in \LRS{} along the $\Gamma-A$ line in Fig.~\ref{Fig1}(d). Therefore, flat-band electrons are not fully localized in known bulk kagome metals.
To achieve tailored flat bands, it has been proposed to either suppress interlayer coupling [$t_\mathrm{out}$ in Fig.~\ref{Fig1}(b)] by isolating kagome layers~\cite{Jovanovic2022-dl}, or to introduce 3D frustrated hopping~\cite{Wakefield2023-zc}.
For the latter approach, 3D flat bands have been predicted for the pyrochlore lattice~\cite{Guo2009-bc}, a 3D analogue of the kagome lattice, and have recently been observed~\cite{Wakefield2023-zc,Huang2024-gp}.
Despite its simplicity, the former approach has not yet been conclusively implemented to enhance electron correlation phenomena and superconductivity, possibly driven by specific phonon modes, on the kagome lattice.

\begin{figure}[htb]
  \begin{center}
    \includegraphics[clip, trim=0cm 0cm 0cm 0cm, width=0.6\linewidth]{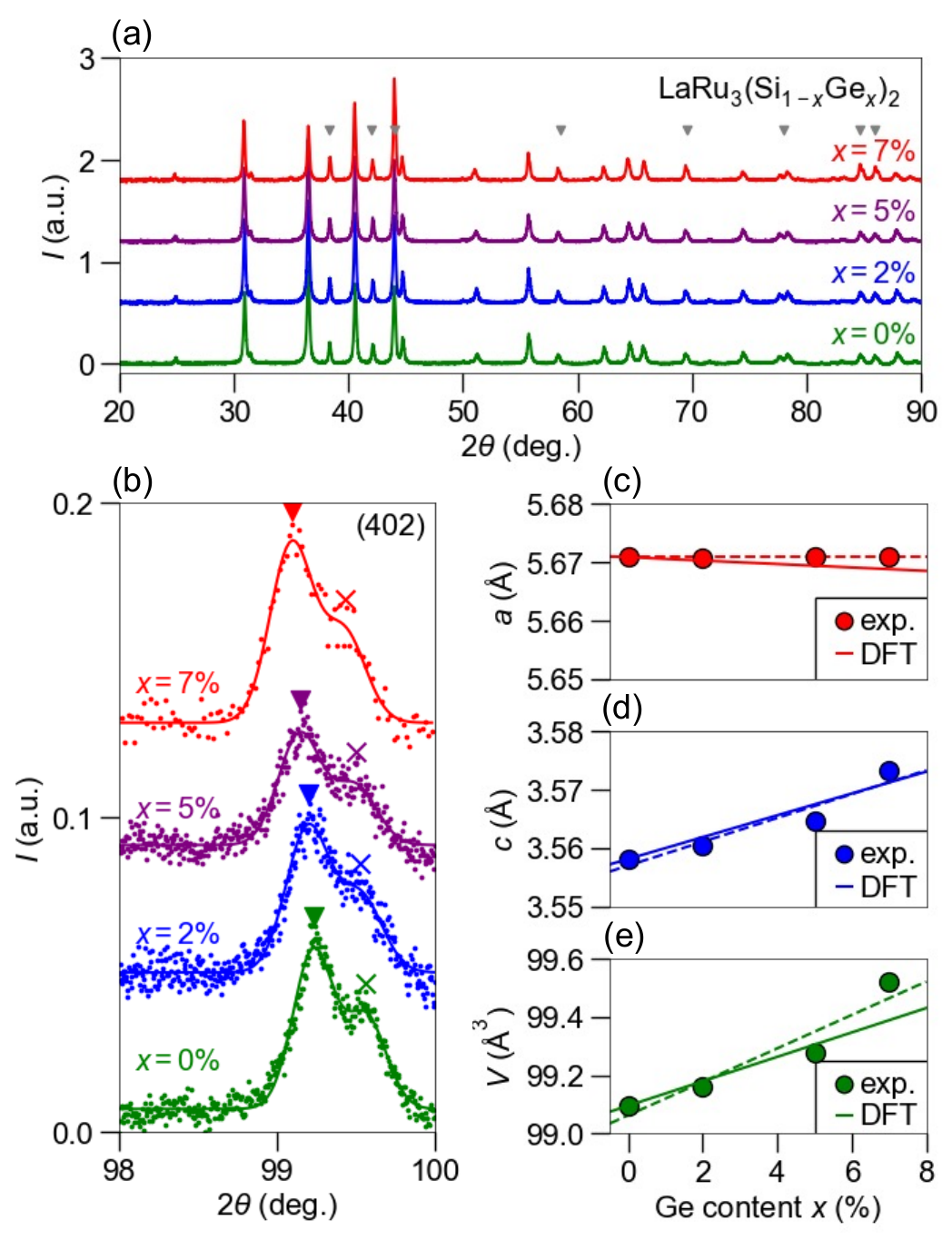}
    \caption[]{(color online). Structural evolution of \LRSG{} (LRSG). (a) Powder X-ray diffraction (XRD) pattern for four compositions of LRSG with normalized intensity.
    Impurity lines from excess Ru metal are indicated by grey triangles. 
    (b) Systematic shift of $(402)$ reflection as a function of $x$, measured using X-rays from Cu-K$\alpha_1$ (triangle) and Cu-K$\alpha_2$ (cross). The lines indicate a double-Gaussian fit to the data.
    Offsets are applied to each compound for clarity in panels (a,b).
    (c-e) Evolution of lattice constants $a$, $c$, and unit cell volume $V$ for LRSG from LeBail analysis of the powder XRD (filled circle)
    and DFT calculations (solid line) in the approximate hexagonal structure. Dotted lines correspond to linear fits to the data. Solid lines are derived from structure optimization by the DFT calculations for \LRS{} and \LRG{}. A constant offset is applied to the DFT data to align with the experimental values at $x = 0$.
    }
    \label{Fig2}
  \end{center}
\end{figure}

Here, we engineer the electronic structure of \LRS{} by substituting Si with the larger volume element Ge. We observe a uniaxial lattice expansion along the $c$-axis and a marked increase of the superconducting critical temperature $T_\mathrm{c}$. An enhancement of the total density of states (DOS) with Ge-doping underpins the observed evolution of $T_\mathrm{c}$, as confirmed by specific heat measurements. Our electronic structure calculations reproduce these behaviors; they suggest the bandwidth of the flat band decreases due to suppression of interlayer hopping $t_\mathrm{out}$ by the anisotropic lattice expansion. The calculations also suggest an effective hole doping through chemical pressure, moving the flat band closer to the Fermi level. The increase in $T_\mathrm{c}$ is broadly consistent with the scenario of mode-selective coupling to kagome phonons in \LRS{}~\cite{Deng2025}. As summarized in Fig.~\ref{Fig1}(e), our results are in stark contrast to previous attempts to tune superconductivity in \LRS{} by chemical substitution at the Ru site, which generally results in a sharp decrease of $T_\mathrm{c}$~\cite{Li2012-fn,Li2016-rh,Chakrabortty2023-ni}.
Crystals are obtained by the melting technique~\cite{SI}. In Fig.~\ref{Fig2}(a), the \LRS{} structure is confirmed by powder X-ray diffraction (XRD). The absence of the LaRu$_2$Si$_2$ phase is confirmed. Figure \ref{Fig2}(c,d) shows the hexagonal lattice constants of \LRSG{} (LRSG) extracted from XRD and their evolution with $x$ based on the LeBail analysis in the $P6/mmm$ structure. This neglects minor effects of symmetry-lowering 
charge order in this family of compounds~\cite{Plokhikh2024-eg}. 
We find a robust value of $a$ and a linear expansion of the out-of-plane lattice constant $c$, which in combination gives a linear volume expansion in Fig.~\ref{Fig2}(e). The latter is consistent with Vegard's law, and in good agreement with the trend derived from structure relaxation calculations of \LRS{} and \LRG{}. 

\begin{figure*}[htb]
  \begin{center}
		\includegraphics[clip, trim=0cm 0cm 0cm 0cm, width=0.95\linewidth]{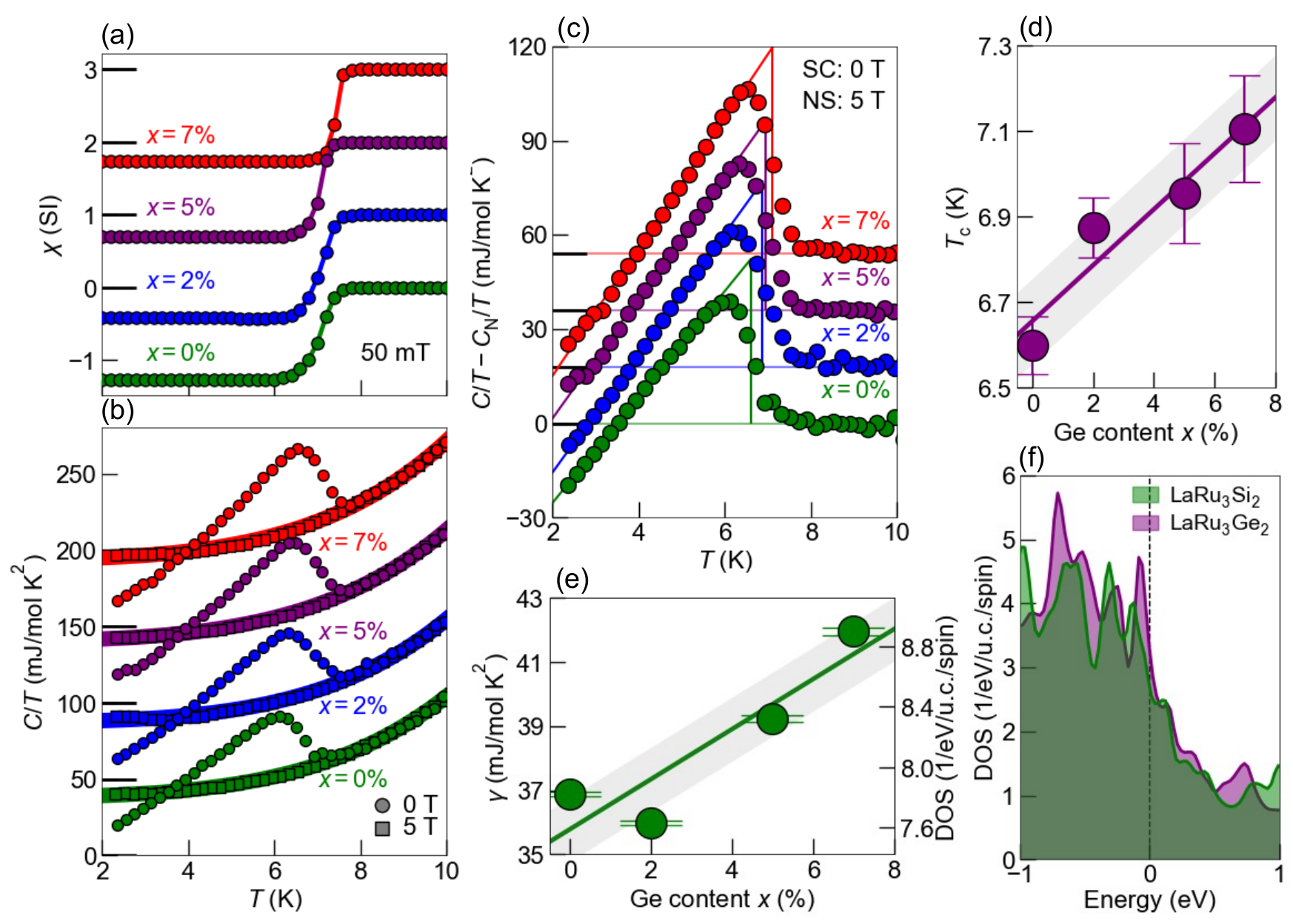}
    \caption[]{(color online). Enhancement of superconductivity by chemical pressure tuning in \LRSG{} (LRSG). (a) Magnetic susceptibility $\chi = M/H$ in the SI unit. The onset of a rapid drop indicates the Meissner effect of a bulk superconducting state; demagnetization correction was applied~\cite{SI}.
    (b) Specific heat $C(T)$ in zero magnetic field (round symbols) and in a field large enough to suppress the superconducting state (square symbols). Lines are a fit to the normal state specific heat $C_\mathrm{N}(T)$ using $C_\mathrm{N}(T) = \gamma T + \beta T^3+ \eta T^5$. 
    (c) After subtraction of 
    $C_\mathrm{N}(T)$, we perform a linear fit for $C(T) - C_\mathrm{N}(T)$ by equal-area construction (solid lines).
    Offsets are applied to each dataset for clarity, with the zero values marked by bold ticks in panels (a-c).
    (d) Evolution of electronic specific heat coefficient $\gamma$ extracted from $C_\mathrm{N}(T)$ as a function of Ge content $x$.
    (e) Superconducting transition temperature $T_\mathrm{c}$. (f) Density of states (DOS) from density functional theory for \LRS{} and (hypothetical) \LRG{}. The DOS at the Fermi level $E_\mathrm{F}$ is $2.9\,\mathrm{/eV/u.c/spin}$ and $3.3\,\mathrm{/eV/u.c./spin}$ for \LRS{} and \LRG{}, respectively. 
    }
    \label{Fig3}
  \end{center}
\end{figure*}

Magnetization and specific heat measurements are performed in a commercial cryostat~\cite{SI}. To confirm the bulk superconducting state in LRSG, we show the magnetic susceptibility $\chi$ in Fig.~\ref{Fig3}(a). Above the transition temperature $T_\mathrm{c}$, $\chi$ is small and positive, consistent with the weak paramagnetic Pauli susceptibility of \LRS{}~\cite{Kishimoto2002-zx}. Strong diamagnetism accompanying the superconducting transition appears as a sharp drop in $\chi$ around $6\sim7\,$K. 

To precisely determine $T_\mathrm{c}$, we measure the specific heat $C(T)$ in zero field and at $5\,$T, which is high enough to entirely suppress the superconductivity [Fig.~\ref{Fig3}(b)]. In the former traces, a sharp kink appears at $T_\mathrm{c}$, the second-order transition into the superconducting state. We model the normal state contribution $C_\mathrm{N}$ by electronic and phonon terms, viz. $C_\mathrm{N}(T) = \gamma T +\beta T^3+ \eta T^5$ [lines in Fig.~\ref{Fig3}(b)].
The higher order phonon contribution $\eta T^5$ is necessary to describe the data
~\cite{Li2011-qv}.
After subtraction of $C_\mathrm{N}(T)$, we fit the superconducting specific heat $C(T) - C_\mathrm{N}(T)$ to the prediction of Bardeen-Cooper-Schrieffer (BCS) theory by equal-area construction in Fig.~\ref{Fig3}(c)~\cite{Tinkham2004-ef},
confirming an enhancement in $T_\mathrm{c}$ with $x$ in Fig.~\ref{Fig3}(d), reaching up to $7.1\,$K for $x = 0.07$.

In Fig.~\ref{Fig3}(e), a positive slope of $\gamma(x)$ is reported. Error analysis and the evolution of $\beta$, from which we calculate $\Theta_\mathrm{D}$, are discussed in the Supplementary Information~\cite{SI}; see also Fig.~\ref{Fig4}(f). Given the formula $\gamma = \pi^2k_\mathrm{B}^2N(E_\mathrm{F})/3$ with $N(E_\mathrm{F})$ being the density of states at the Fermi energy $E_\mathrm{F}$, the enhancement of $\gamma$ corresponds to an increase of $N(E_\mathrm{F})$ [Fig.~\ref{Fig3}(e)].
The trend is reproduced by our density functional theory (DFT) calculations, where $N(E_\mathrm{F})$ increases from $2.9\, \mathrm{/eV/u.c/spin}$ to $3.3\,\mathrm{/eV/u.c/spin}$ when going from \LRS{} to \LRG{}, c.f. Fig.~\ref{Fig3}(f). Calculations at an intermediate $x$ value are discussed in Ref.~\cite{SI}. We now take a closer look at the electronic bands to understand the behavior of $N(E_\mathrm{F},x)$.
The projected band structures for \LRG{} and \LRS{} are shown in Fig.~\ref{Fig4}(a),(b). Focusing on the $\Gamma-A$ line, which is parallel to the $k_z$-axis,
we find that substitution of Si with Ge decreases the bandwidth of the flat band at $E_\mathrm{F}$; the uniaxial lattice expansion along the $c$-axis by Ge doping suppresses interlayer coupling between kagome layers.

In the partial density of states (PDOS) of \LRG{}, Fig.~\ref{Fig4}(c),(d), the contribution of $4d_{z^2}$ grows weaker while the PDOS peak of $4d_{x^2-y^2}$ sharpens, corresponding to a narrowing of the flat band through chemical pressure tuning. Moreover, the calculations indicate a slight shift of the PDOS peak towards $E_{\mathrm{F}}$. From the viewpoint of superconductivity, Ge doping has the favorable effect of reducing the bandwidth, but it also helps to drive $E_\mathrm{F}$ into the fractionally filled flat band. Note that recent theory work predicts an enhancement of $T_\mathrm{c}$ by hole doping \LRS{}~\cite{Deng2025}.

To discuss these results, we consider the McMillan formalism for BCS,
\begin{equation}
\label{eq:McMillan}
T_\mathrm{c} = \frac{\left<\omega\right>}{1.20} \exp\left( \frac{-1.04(1+\lambda)}{\lambda - \mu^*(1+0.62\lambda)} \right),
\end{equation}
where $\left<\omega\right>$ is a characteristic phonon frequency after weighting by the electron-phonon interaction~\cite{SI} and $\mu^*$ is the Coulomb pseudopotential~\cite{McMillan1968-am}.
The electron-phonon coupling constant $\lambda$ can be expressed as
\begin{equation}
\label{eq:lambda}
    \lambda = 2\frac{N(E_\mathrm{F})}{N}\frac{\hbar\langle g^2\rangle}{\hbar^2\langle \omega^2\rangle},
\end{equation}
where $N$ and $\langle g^2\rangle$ represent the number of sites and the weighted phonon linewidth
~\cite{McMillan1968-am}. For \LRS{} we expect $\lambda<1$ based on (i) recent \textit{ab-initio} modeling, which predicts $\lambda \approx 0.8$, and (ii) $\Delta C/(\gamma T_\mathrm{c})$ in Fig.~\ref{Fig4}(h), which is characteristic of moderate coupling strength. 

According to recent theory, $d_{x^2-y^2}$ flat bands have a mode-selective electron-phonon coupling to the $B_{3u}$ kagome phonon~\cite{Deng2025}. This low-energy mode corresponds to a planar oscillation into and out of the large hexagonal plaquette in the kagome lattice, illustrated in Fig.~\ref{Fig4}(e). In the scenario of Deng \textit{et al.}~\cite{Deng2025}, the widely used~\cite{Dynes1972} 
\begin{equation}
\label{eq-dynes}
\left<\omega\right> \approx0.828\, \Theta_\mathrm{D}
\end{equation} 
cannot be valid for \LRSG{}: the Debye temperature is an average over the phonon spectrum $F(\omega)$ that is not weighted by the electron-phonon coupling $\alpha^2(\omega)$, unlike $\left<\omega\right>= \int \mathrm{d}\omega \alpha^2(\omega)F(\omega)/\left[\int \mathrm{d}\omega \alpha^2(\omega)F(\omega)/\omega\right]$~\cite{McMillan1968-am,Dynes1972,Allen1975}. For the sake of argument, we apply Eq.~(\ref{eq-dynes}) to \LRS{} with $N(E_\mathrm{F})$ from $\gamma$ and show $N\left<\omega^2\right>/\left<g^2\right>$ from Eq.~(\ref{eq:lambda}) in Fig.~\ref{Fig4}(g). While the measured $\Theta_\mathrm{D}$ softens with $x$, the latter quantity grows in equal measure. This lack of agreement between Fig.~\ref{Fig4}(f) and (g) suggests the \textit{mode independent} approximation for $\alpha^2(\omega)$, Eq.~(\ref{eq-dynes}), is not suitable in \LRS{}. However, BCS theory and Eq.~(\ref{eq:McMillan}) remain valid for this kagome metal, even with mode-selective electron-phonon coupling; $N(E_\mathrm{F})$ and the reported flattening of the $4d_{x^2-y^2}$ band, not changes in the phonon spectrum, predominantly drive a Ge-induced enhancement of superconductivity~\cite{SI}.

In summary, we control superconductivity in \LRS{} by chemical pressure and use specific heat measurements -- which is not accessible for recent studies of enhanced superconductivity in \LRS{} under high external pressure~\cite{Ma2024-rw,Li2025-xu} -- to track the electronic density of states and phononic Debye temperature. 
Although chemical pressure tuning has been implemented in a number of superconductors~\cite{SI}, simultaneous reports of $T_\mathrm{c}$ and $\gamma$ are less common, and none such has targeted kagome metals. 
Tuning the flat band by uniaxial strain, using large single crystals, remains a challenge for future research. \\

\textbf{Acknowledgements}
We thank T.-h. Arima for enlightening discussions. This work was supported by JSPS KAKENHI Grants Nos. JP21K13877, JP22H04463,  JP22K20348,  JP23H05431, 23K13057, 24H01607, and 24H01604, as well as JST CREST Grant Nos. JPMJCR1874 and JPMJCR20T1 (Japan) and JST FOREST Grant No. JPMJFR2238 (Japan). This work was supported by Japan Science and Technology Agency (JST) as part of Adopting Sustainable Partnerships for Innovative Research Ecosystem (ASPIRE), Grant Number JPMJAP2426. R.Y. acknowledges support from the US National Science Foundation (NSF) Grant Number 2201516 under Accelnet program of Office of International Science and Engineering (OISE). Work at Princeton was supported by NSF, grant OAC-2118310, and by NSF through the Princeton Center for Complex Materials, a Materials Research Science and Engineering Center DMR-2011750.
\bibliography{LaRu3Si2}

\begin{thebibliography}{59}
\bibitem [{foo()}]{footnote1}%
  \BibitemOpen
  \href@noop {} {}\bibinfo {note} {{{{{Different reports in the literature show slightly different values of $T_\mathrm{c}$ of pure \LRS{}. We shift the $T_\mathrm{c}(x)$ so that $T_\mathrm{c}(0)$ becomes 6.6 K}}}}}\BibitemShut {NoStop}%
\end{thebibliography}

\clearpage

\begin{figure*}[htb]
  \begin{center}
		\includegraphics[clip, trim=0cm 0cm 0cm 0cm, width=0.95\linewidth]{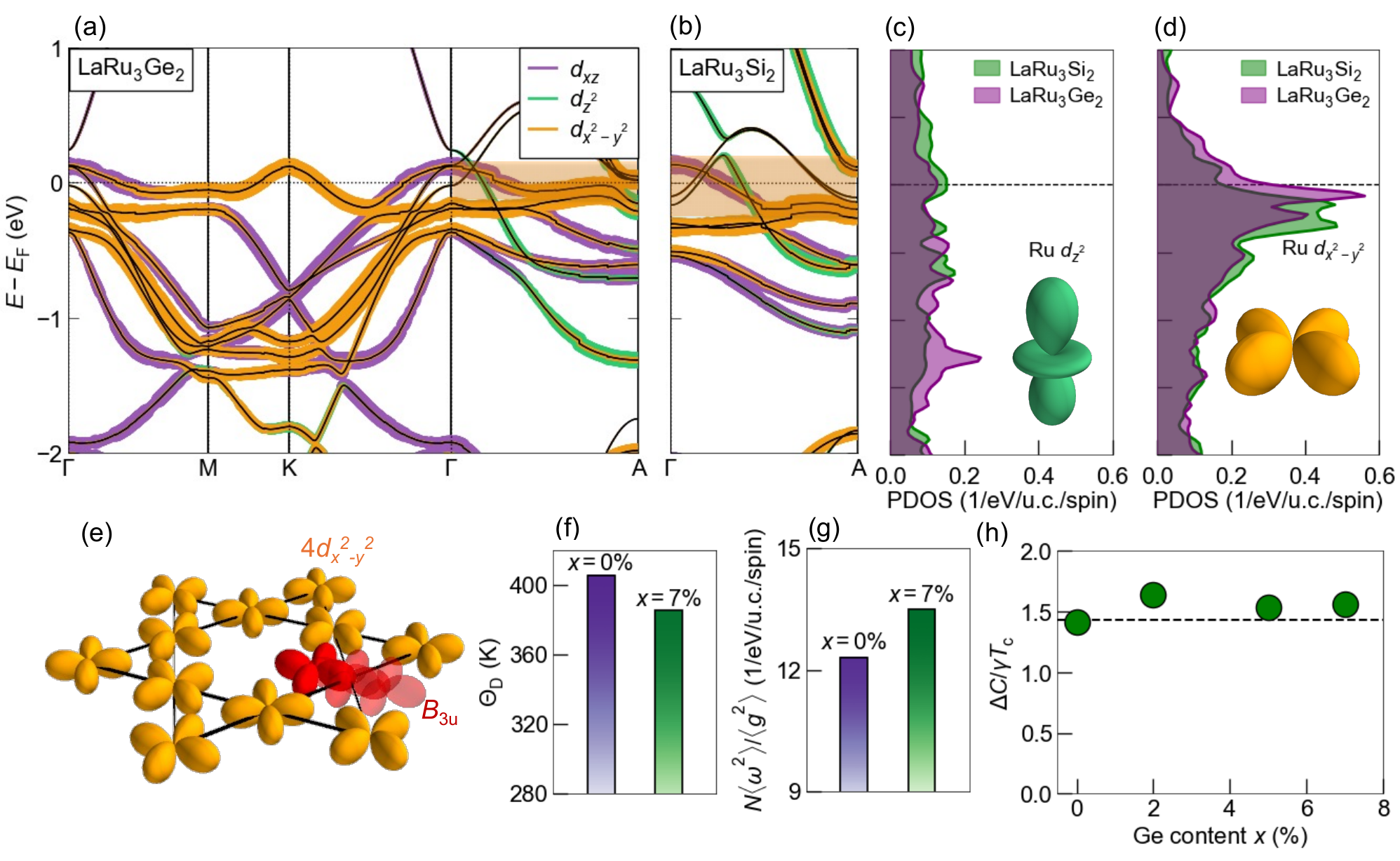}
    \caption[]{(color online). Electronic structure of \LRG{} in comparison with \LRS{}.
    (a) Projected band structure of \LRG{} in the $k_x-k_y$ plane and along the $\Gamma-A$ line.
    Bands that are flat in the $k_x-k_y$ plane are highlighted in orange along the path parallel to the $k_z$-axis. 
    (b) Projected band structure of \LRS{} along the $\Gamma-A$ line. Bandwidth of the flat band along this path is wider for \LRS{}.
    (c, d) Partial density of states (PDOS) for Ru-$4d_{z^2}$ and Ru-$4d_{x^2-y^2}$ states of \LRS{} and \LRG{}. The contributions of $d_{z^2}$ and $d_{x^2-y^2}$ at $E_\mathrm{F}$ are suppressed and enhanced by replacing Si with Ge, respectively. (e) Illustration of $d_{x^2-y^2}$ orbitals on the lattice and kagome phonon mode $B_{3u}$, discussed by Deng \textit{et al.}~\cite{Deng2025}. $B_{3u}$ corresponds to vibration into and out of the central hexagon of the kagome pattern (red highlight, with shading indicating time dependence). (f,g) Evolution of $\Theta_\mathrm{D}$ extracted from the specific heat measurement and $\lambda/N(E_\mathrm{F})$ from Eq.~(\ref{eq:lambda}). (h) Specific heat jump $\Delta C$ at $T_\mathrm{c}$, normalized by electronic specific heat just above the transition temperature. Dashed line: moderate coupling BCS value. 
    }
    \label{Fig4}
  \end{center}
\end{figure*}
\begin{widetext}
\clearpage
\newpage
\section*{Supplementary Information}

\renewcommand\thefigure{S\arabic{figure}} 
\setcounter{figure}{0}

\renewcommand\theequation{S\arabic{equation}}
\setcounter{equation}{0}

\renewcommand\thesection{S \Roman{section}} 
\setcounter{section}{0}

\renewcommand\thetable{S\arabic{table}} 
\setcounter{table}{0}
\section{Crystal growth}

Crystals of \LRSG{} (LRSG) are prepared by arc melting in an argon atmosphere after careful evacuation to $10^{-3}\,$Pa. The ingot is turned and remelted at least three times, and weight loss is typically around $0.5\,\%$. To stabilize the \LRS{} structure over the competing phase LaRu$_2$Si$_2$, excess Ru metal is added to the starting composition~\cite{Barz1980-ep,Vandenberg1980-he,Kishimoto2002-zx,Li2011-qv,Mielke2021-ha}. The solution limit of Ge is found to be $x = 0.07$, and the required amount of Ru increases when approaching this value: We use a melt of starting composition LaRu$_{4.5}$(Si$_{1-x}$Ge$_{x}$)$_2$ for $x = 0, 0.02, 0.05, $ and LaRu$_{5}$(Si$_{1-x}$Ge$_{x}$)$_2$ for $x = 0.07$.

\section{Powder X-ray diffraction}
We perform powder X-ray diffraction (XRD) to confirm the formation of LRSG as well as the absence of the competing LaRu$_2$Si$_2$. The data include reflections from nonmagnetic, elemental Ru, because Ru is added to the melt to destabilize the chemically stable LaRu$_2$Si$_2$ phase. 


\section{Density functional calculation}
First-principles calculations are carried out using the WIEN2k software package, which employs the full-potential linearized augmented plane wave (FP-LAPW) method with local orbitals~\cite{Blaha2020-ns}. The exchange-correlation energy is treated within the Generalized Gradient Approximation (GGA) using the functional formulated by Perdew, Burke, and Ernzerhof (PBE)~\cite{Perdew1996-ih} in the $P6/mmm$ structure. We employ a uniform grid of 4096 $k$-points to ensure accurate sampling of the Brillouin zone and relax the lattice constants $a$ and $c$.
This yields $a=5.8260 \mathrm{\AA},\ c = 3.5666 \mathrm{\AA}$ for \LRS{} and $a = 5.7967 \mathrm{\AA},\ c = 3.7538 \mathrm{\AA}$ for hypothetical \LRG{}. We calculate the band structures and the density of states in the presence of spin-orbit coupling (SOC).
For the density of states (DOS) and partial density of states (PDOS), we divide the calculated values by $2$ to express them in units of $\mathrm{eV}^{-1}$ per unit cell (u.c.) per spin. A previous report on the band structure of \LRS{} assumes the $P6_3/m$ structure with two time larger unit cell~\cite{Mielke2021-ha}. However, a recent study shows that the proper structure is orthorhombic~\cite{Plokhikh2024-eg}. Due to the different structure and band folding in the $P6_3/m$ space group, slight differences can be seen in our calculation results compared to the previous study. On the other hand, our calculations are largely consistent with recent theoretical work on \LRS{}~\cite{Deng2025}. For our calculations, we ignore the structural transition from $P6/mmm$ to $Cccm$ and subsequent CDW transitions~\cite{Plokhikh2024-eg}. Recent theoretical work shows that the flat band persists in the orthorhombic $Cccm$ phase~\cite{Deng2025}. 


Fig.~\ref{Fig4} shows the PDOS for the Ru $4d$ orbitals of \LRS{} and \LRG{}. We provide projected band structures in the full Brillouin zone for \LRS{} and \LRG{} in Fig.~\ref{SI_Fig4}.

To check the impact of spin-orbit coupling (SOC), we compare band structure and DOS with and without SOC in Fig.~\ref{SI_Fig5}. For both cases, the flat band is present at the Fermi energy, and \LRG{} has a larger DOS than \LRS{}.

In addition to \LRS{} and \LRG{}, we illustrate in Fig.~\ref{SI_Fig6} the calculations performed for LRSG with $x=8\, \%$ by the virtual crystal approximation (VCA)~\cite{Bellaiche2000-fe} implemented in VASP~\cite{Kresse1996-uh}. The DOS systematically increases around the Fermi level upon doping Ge [Fig.~\ref{SI_Fig6}(a)].

\begin{figure}[htb]
  \begin{center}
    \includegraphics[clip, trim=0cm 0cm 0cm 0cm, width=1.\linewidth]{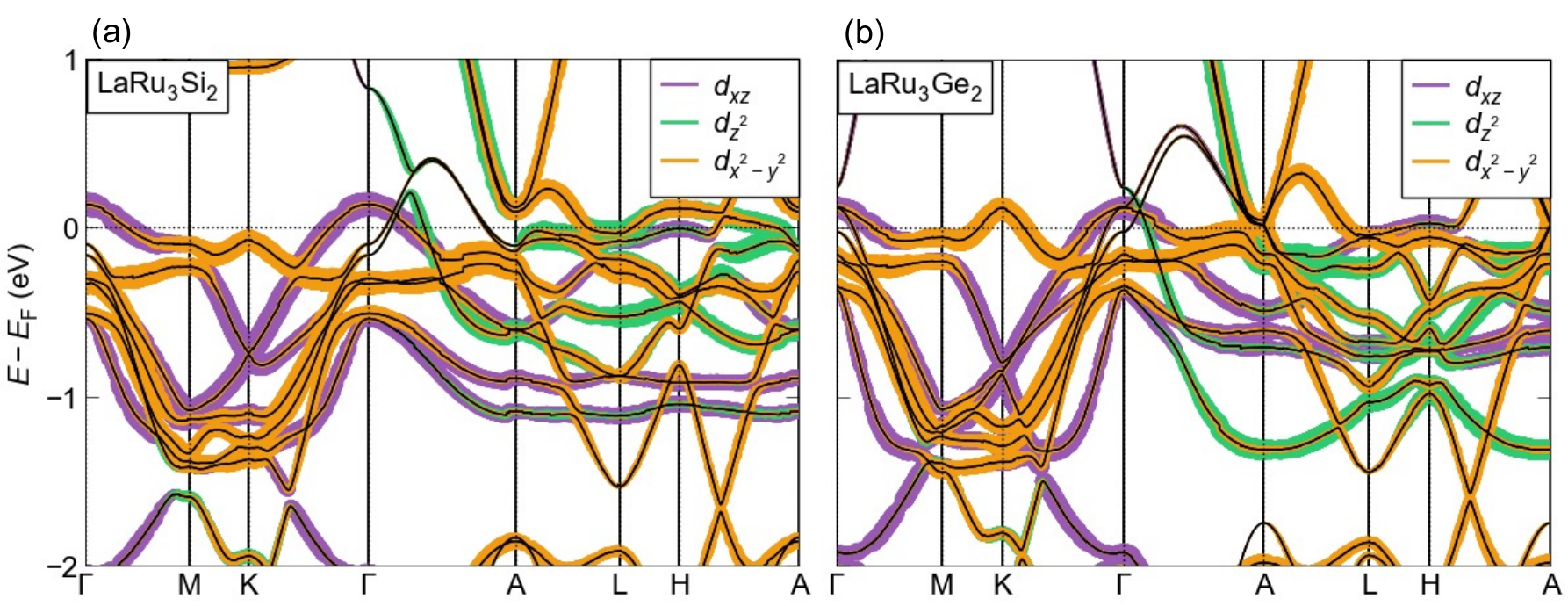}
    \caption[]{(color online). Projected band structure of \LRS{} (a) and \LRG{} (b) in the full Brillouin zone.}
    \label{SI_Fig4}
  \end{center}
\end{figure}
\begin{figure}[htb]
  \begin{center}
    \includegraphics[clip, trim=0cm 0cm 0cm 0cm, width=1.\linewidth]{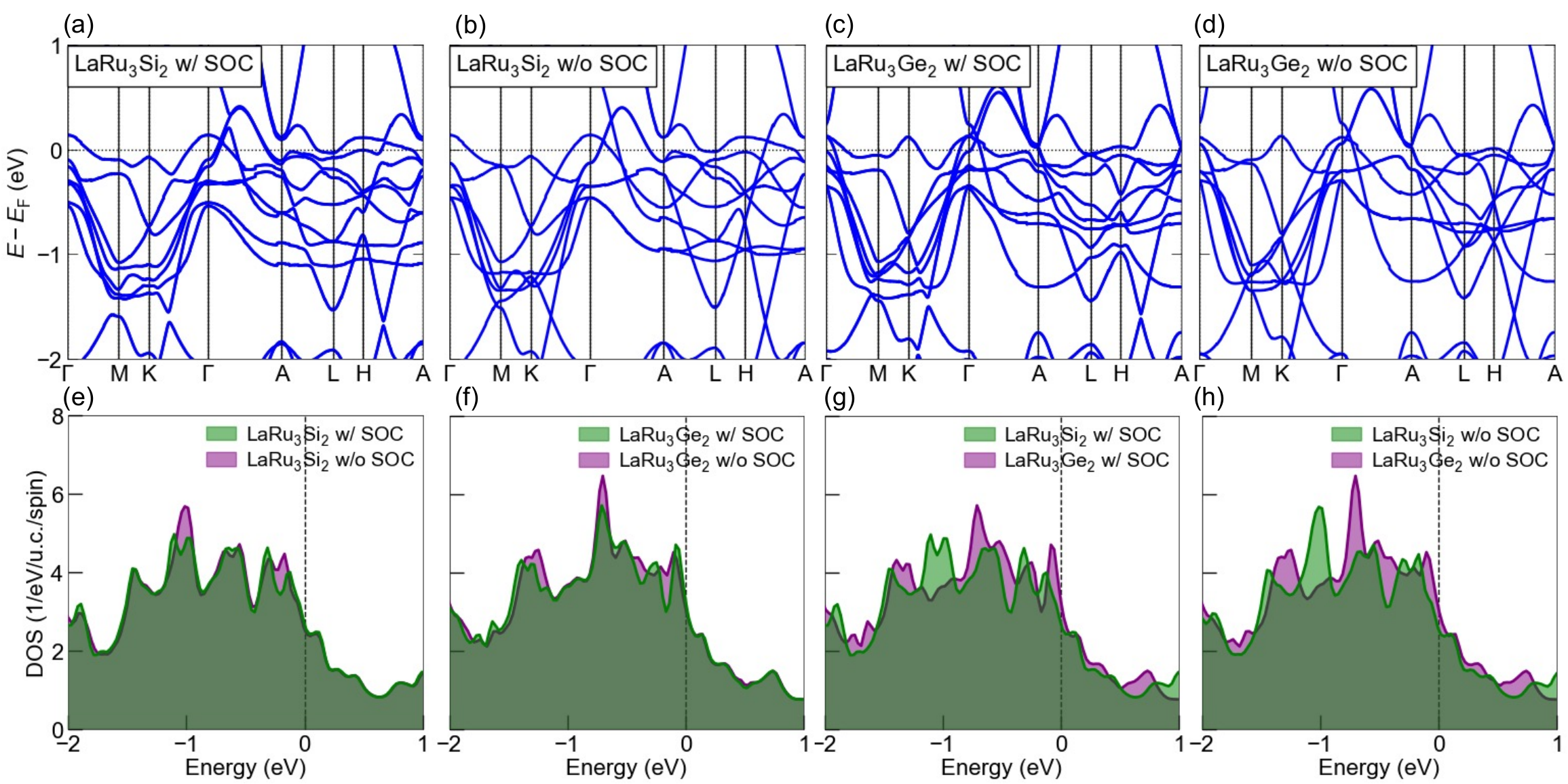}
    \caption[]{(color online). Comparison of \textit{ab initio} calculations with and without SOC.
    (a),(b) Band structure for \LRS{} with and without spin-orbit coupling (SOC). (c),(d) Band structure for \LRG{} with and without SOC. (e) Density of states (DOS) for \LRS{} with and without SOC. (f) DOS for \LRG{} with and without SOC. (g) Comparison of \LRS{} and \LRG{} with SOC. (h) Comparison of \LRS{} and \LRG{} without SOC.
    }
    \label{SI_Fig5}
  \end{center}
\end{figure}
\begin{figure}[htb]
  \begin{center}
    \includegraphics[clip, trim=0cm 0cm 0cm 0cm, width=1\linewidth]{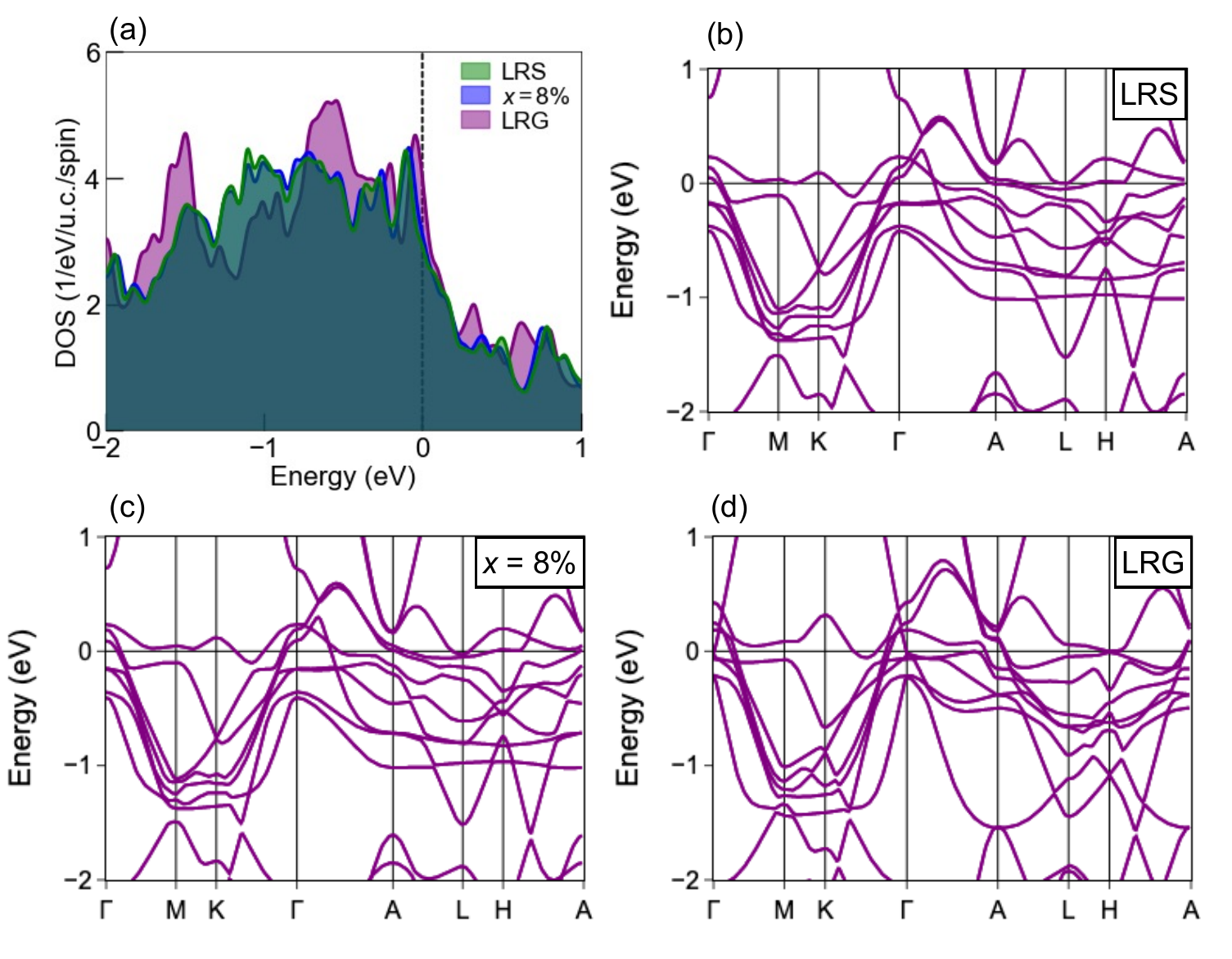}
    \caption[]{(color online). First-principle calculations for \LRSG{} (LRSG) with $x=0,\, 0.08,\, 1$ performed in VASP. For $x=0.08$, virtual crystal approximation is employed.
    (a) Comparison of the DOS. (b)-(d) Band structure for \LRS{} (LRS), LRSG ($x=0.08$), and \LRG{} (LRG).
    }
    \label{SI_Fig6}
  \end{center}
\end{figure}

\section{Magnetic susceptibility measurement}

We perform magnetization measurements to detect the superconducting Meissner effect in a low magnetic field of $\mu_0 H= 0.05\,$T. Samples of mass $\sim 100\,$mg are mounted in a Quantum Design
Magnetic Properties Measurement System (MPMS). We correct for the demagnetization effect in $\chi$ by approximating our roughly shaped crystals as a sphere, which has a demagnetization factor $N = 1/3$. The intrinsic susceptibility is then obtained using the relation:
$\chi = \chi_{\mathrm{exp}}/(1 - N\chi_{\mathrm{exp}})$.
Slightly smaller values than the ideal $\chi = -1$ (perfect Meissner shielding) may arise due to deviations from the spherical approximation.

\section{Specific heat measurement}
We measure specific heat by the relaxation technique in a Quantum Design Physical Properties Measurement System (PPMS).
For each measurement, the addenda (with Apiezon N grease mounted on the stage) are measured before setting the sample for the main measurement.
To eliminate contributions from non-magnetic impurity Ru metal, we subtract $\gamma_\mathrm{Ru} = 2.8 \, \mathrm{mJ/mol K^2}$ and $\beta_\mathrm{Ru} = 0.068 \, \mathrm{mJ/mol K^4}$ from the total $\gamma$ and $\beta$ values~\cite{Li2011-qv,Reese1970-fi}.
For each value, we multiply the ratio of Ru impurity $y/3$, where $y$ represents the ratio of excess Ru, namely $y = 1.5$ for $x = 0, 0.02, 0.05$ and $y = 2$ for $x = 0.07$.
We summarize $\gamma$ and $\beta$ values before and after the subtraction of the Ru contribution in Table \ref{tab:gamma_beta_values}.

Errors for the fitting parameters $\beta$ and $\gamma$ are estimated by the bootstrap method. This allows us to evaluate errors or the stability of fitting while incorporating systematic measurement errors. The procedure is as follows: We first fit the data and calculate residuals. We then compute the total measurement error for each data point by performing the relaxation method measurement 3 times at each temperature and then calculating the mean and standard deviation $\sigma(T)$. Next, we randomly sample values from zero to $\sigma(T)$, add them to the experimental data and obtain fit parameters for this perturbed data. This is repeated 100 times. We take an average of these trials as a central value for $\beta$ and $\gamma$; we also use its standard deviation as a fit error for these parameters.  Other errors, such as sample mass and batch dependence, are ignored as we precisely measured the mass and uniformly melted all samples.

\section{Electronic and phononic specific heat}
We estimate the density of states (DOS) at the Fermi energy, $N(E_\mathrm{F})$, using the relation  
\begin{equation}
    N(E_\mathrm{F}) = \frac{3\gamma}{\pi^2 k_\mathrm{B}^2},
\end{equation}
which has units of $\mathrm{J}^{-1}$. We convert this to $\mathrm{eV}^{-1}$ per unit cell (u.c.). Furthermore, since $\gamma$ is calculated assuming a spinless electronic structure, we divide by 2 to obtain $N(E_\mathrm{F})$ in units of $\mathrm{eV}^{-1}$ per unit cell per spin~\cite{Tinkham2004-ef,Kriener2022-vx}. The errors of $N(E_\mathrm{F})$ in Fig.~\ref{Fig3}
(e) are estimated by the error propagation of $\gamma$. In Table ~\ref{tab:gamma_beta_values}, we also present the Debye temperature $\Theta_\mathrm{D}$ calculated from $\beta$ as
\begin{equation}
    \Theta_\mathrm{D} = \left(\frac{12\pi^4Nk_\mathrm{B}}{5\beta}\right).
\end{equation}

In the Main Text, we discuss the observed reduction of $\Theta_D$ which is summarized in Table \ref{tab:gamma_beta_values}. Assuming Eq.~(\ref{eq-dynes}), of the Main Text, this may be understood to imply a softening of $\left<\omega\right>$ with $x$ for the phonon modes relevant to the superconducting $T_\mathrm{c}$. However, softening appears unreasonable considering the dominant role of the $B_{3u}$ phonon in DFT~\cite{Deng2025}, its nature as a vibration in the hexagonal $ab$ plane, and the lack of a lattice expansion along $a$ in Fig.~\ref{Fig2}(c). The observed softening of $\Theta_\mathrm{D}$ with $x$ should then be attributed to the elastic coupling between kagome layers, which is irrelevant to the superconductivity~\cite{Deng2025}.\\

\section{Specific heat jump at $T_\mathrm{c}$}
For the superconducting state, the specific heat can be fitted by an analytical expression within the framework of the BCS theory.
Here, we approximate the temperature dependence to be linear below $T_\mathrm{c}$, which is valid for our temperature range ($>2\,$K)~\cite{Gui2022-mq}.
This method allows us to precisely determine $T_\mathrm{c}$ by the equal-area construction, which is crucial for tracking the evolution of $T_\mathrm{c}$ in our crystals:
We first numerically calculate the area under the curve of $\Delta C/T = C/T - C_\mathrm{N}/T$ below $8\,$K, denoted as $S$ [Fig.~\ref{SI_Fig7}(a)].
We then fit $\Delta C/T$ below $6\,$K by a linear function $\Delta C/T = a T + b$. Imposing entropy conservation, the area under the curve is thus given by
\begin{equation}
    S = \frac{1}{2}(T_\mathrm{c} - T_0)(a T_\mathrm{c} + b) - \frac{1}{2}(T_0 - T_\mathrm{min})|a T_\mathrm{min} + b|,
\end{equation}
where $T_0 = -b/a$ is the intercept of the linear fit, and $T_\mathrm{min}$ is the lowest temperature at which data are taken.
The first term corresponds to the area for $T_0 \leq T \leq T_\mathrm{c}$ and the second to that for $T_\mathrm{min} \leq T \leq T_\mathrm{c}$, as illustrated in Fig.~\ref{SI_Fig7}(b).
Solving this equation, we analytically determine the transition temperature as
\begin{equation}
  T_\mathrm{c} = \frac{-b + \sqrt{b^2 + 2aS +a^2T_\mathrm{min}^2 + 2 abT_\mathrm{min}}}{a}.
\end{equation}
Furthermore, we obtain the specific heat jump at $T_\mathrm{c}$ as
\begin{equation}
    \Delta C(T_\mathrm{c})/T_\mathrm{c} = aT_\mathrm{c} + b.
\end{equation}
The errors of $a$ and $b$ are evaluated by the same method as for $\beta$, $\gamma$, and $T_\mathrm{c}$ described above and shown in Fig.~\ref{Fig3}. Error propagation was carefully considered.

We provide the calculated ratio $\Delta C(T_\mathrm{c})/\gamma T_\mathrm{c}$ in Table \ref{tab:gamma_beta_values}. A slight deviation from the BCS value of $1.43$ may suggest a moderate superconducting coupling strength. This behavior is consistent with a previous report~\cite{Chakrabortty2023-ni}. Given that the Sommerfeld coefficient of pure Ru is small ($\gamma_\mathrm{Ru} = 2.8\,\mathrm{mJ/mol\, K^2}$), the residual specific heat contribution from the Ru impurity is expected to be negligible. \\

\section{Electron-phonon coupling strength and bare density of states}
Using Eq.~(\ref{eq:McMillan}) and Eq.~(\ref{eq-dynes}) of the Main Text, we estimate the electron-phonon coupling constant $\lambda$.
Setting $\mu^*=0.13$, we estimate $\lambda$ to be $0.63 \sim 0.66$, as summarized in Table \ref{tab:gamma_beta_values}. The enhanced electron-phonon coupling constant $\lambda > 0.5$ suggests a moderately coupled superconductor. This is consistent with $\Delta C(T_\mathrm{c}) / \gamma T_\mathrm{c} > 1.43$ in Fig.~\ref{Fig4}(h).

With $\lambda$ determined, we compute the bare density of states, $N_\mathrm{bare}(E_\mathrm{F})$, using  
\begin{equation}
    N_\mathrm{bare}(E_\mathrm{F}) = \frac{N(E_\mathrm{F})}{1+\lambda},
\end{equation}
which allows for comparison to band structure calculations. Here, $N_\mathrm{bare}(E_\mathrm{F})$ has the same unit as $N(E_\mathrm{F})$ because $\lambda$ is dimensionless.
This yields approximately $N_\mathrm{bare}(E_\mathrm{F}) = 5\,\mathrm{/eV/u.c/spin}$ and reasonably corresponds to the calculated DOS at $E=-0.1\,\mathrm{eV}$ in Fig.~\ref{Fig3}(f).\\

\section{Strength of doping effect on $T_\mathrm{c}$}
Table~\ref{tab:Tc_doping} shows examples of superconductors with isovalent doping~\cite{Xie2023-ld,Kriener2018-vg,Ozaki2013-ym,Grassie1972-xt,Liu2021-tz,Li2022-uy,Liu2024-hf}. Since the resolution limit $x_0$ -- the maximum doping value studied -- is different for various systems, we compare the enhancement of $T_\mathrm{c}$, $\Delta T_\mathrm{c}$, divided by $x_0$. 
Crucially, our study systematically traces the evolution of $N(E_\mathrm{F})$ and compares the results to first-principle calculations, revealing the origin of the enhanced $T_\mathrm{c}$. 
We note that 4$H_\mathrm{b}$-Ta(S$_{1-x}$Se$_x$)$_2$ shows a decreasing $T_\mathrm{c}$ above $x_0=30\, \%$ and CsV$_3$Sb$_5$ is doped at the transition metal site (V), in contrast to other systems.

\begin{figure}[htb]
  \begin{center}
    \includegraphics[clip, trim=0cm 0cm 0cm 0cm, width=1\linewidth]{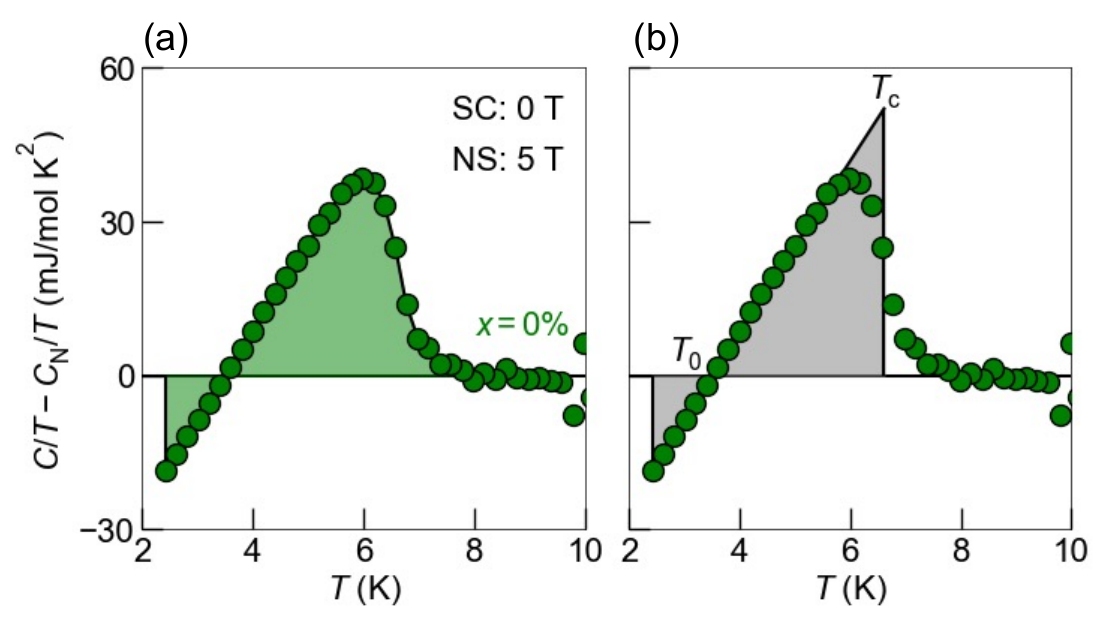}
    \caption[]{(color online). Fitting of superconducting specific heat $\Delta C = C-C_\mathrm{N}$ and equal-area construction. As a representative example, we show the procedure for \LRS{}. (a) The area under the curve (green) is calculated numerically. (b) $\Delta C/T$ is fitted by a
    linear approximation of the BCS expression. $T_\mathrm{c}$ is identified such that the gray areas equal the green area in panel (a). $T_\mathrm{0}$ denotes the intercept of the fit.
    }
    \label{SI_Fig7}
  \end{center}
\end{figure}
\begin{table}[h!]
    \centering
    \caption{Properties of \LRSG{} probed by specific heat measurements. For the electron and phonon contributions $\gamma$ and $\beta$,
    values before and after subtracting a subtle contribution from Ru impurity in \LRSG{} + $y$Ru are given.
    $\Delta C(T_\mathrm{c})$ denotes the specific heat jump at $T_\mathrm{c}$, and $\Delta C(T_\mathrm{c})/\gamma T_\mathrm{c}=1.43$ holds for weakly coupled superconductors.
    $N(E_\mathrm{F})$ and $\lambda$ denote the DOS and the electron-phonon coupling constant, respectively.
    }
    \begin{tabular}{|c|c|c|c|c|}
        \hline
        Ge content $x$ & 0 & 0.02 & 0.05 & 0.07 \\
        \hline
        $\gamma\,\,\ \mathrm{(mJ/mol K^2)}$ & 38.3 & 37.4 & 40.7 & 43.8 \\
        $\gamma' = \gamma - y/3\cdot\gamma_\mathrm{Ru}\,\, \mathrm{(mJ/mol K^2)}$ & 36.9 & 36.0 & 39.3 & 42.0 \\
        $\beta\,\, \mathrm{(mJ/mol K^4)}$ & 0.21 & 0.19 & 0.25 & 0.25 \\
        $\beta'=\beta - y/3\cdot\beta_\mathrm{Ru}\,\, \mathrm{(mJ/mol K^4)}$ & 0.18 & 0.15 & 0.21 & 0.20 \\
        $\Theta_\mathrm{D}$ (K) & 405(6) & 423(20) & 379(6) & 384(9) \\ 
        $\Delta C(T_\mathrm{c})/\gamma' T_c$ & 1.42 & 1.64 & 1.54 & 1.57 \\
        $N(E_\mathrm{F})\,\,\mathrm{(1/eV/u.c/spin)}$
        &7.82& 7.64 &8.33 & 8.90 \\
        $\lambda$ & 0.635(6) & 0.635(6) & 0.659(7) & 0.661(6) \\
        $N(E_\mathrm{F})/(1+\lambda)\,\,\mathrm{(1/eV/u.c/spin)}$ & 4.78 & 4.68 & 5.03 & 5.36 \\
        \hline
    \end{tabular}
    \label{tab:gamma_beta_values}
\end{table}
\begin{table}[h]
    \centering
    \begin{tabular}{|c|c|c|c|c|c|c|c|c|}
        \hline
        Material & $T_\mathrm{c}(0)\,$(K) & $x_0\ $(\%)& $T_\mathrm{c}(x_0)\ $(K) & $\Delta T_\mathrm{c}\ $(\%) & $\Delta T_\mathrm{c}/x_0$ & $N(E_\mathrm{F})_\mathrm{exp}$& $N(E_\mathrm{F})_\mathrm{DFT}$ & reference\\
        \hline
        \hline
        InTe$_{1-x}$Se$_x$ & $2.6$ & $30$ & $5.0$ & $92$ & $3$ & $\times$ & $\times$ & ~\cite{Kriener2018-vg}\\
        1T-Pd(Te$_{1-x}$Se$_x$) & $1.64$ & $50$ & $2.7$ & $67$ & $1$ & $\bigcirc$ & $\times$ & ~\cite{Liu2021-tz}\\
        Sn$_{1-x}$Ge$_x$Te & $2.1$ & $1.2$ & $1.3$ & $-38$ & $-32$ & $\times$ & $\times$ & ~\cite{Grassie1972-xt} \\
        \hline
        Sn$_{1-x}$Pb$_x$Te & $2.1$ & $12$ & $1.9$ & $-10$ & $-0.8$ & $\times$ & $\times$ &~\cite{Grassie1972-xt} \\
        K$_{0.8}$Fe$_2$(Se$_{1-x}$Te$_x$)$_2$ & $31$ & $25$ & $14.9$ & $-52$ & $-2$ & $\times$ & $\times$ & ~\cite{Ozaki2013-ym}\\
        4$H_\mathrm{b}$-Ta(S$_{1-x}$Se$_x$)$_2$ & $2.8$ & $30$ & $4.0$ & $43$ & $1$ & $\times$ & $\times$ & ~\cite{Xie2023-ld}\\
        Cs(V$_{1-x}$Nb$_x$)$_3$Sb$_5$ & $2.9$ & $7$ & $4.5$ & $55$ & $8$ &$\times$ &$\times$ & ~\cite{Li2022-uy}\\
        Cs(V$_{1-x}$Ta$_x$)$_3$Sb$_5$ & $3.1$ & $16$ & $5.3$ & $71$ & $4$ & $\times$ & $\times$ & ~\cite{Liu2024-hf}\\
        LaRu$_3$(Si$_{1-x}$Ge$_x$)$_2$ & $6.6$ & $7$ & $7.1$ & $7$ & $1$ & $\bigcirc$ & $\bigcirc$ & this work\\
        \hline
    \end{tabular}
    \caption{Superconducting transition temperatures $T_\mathrm{c}$ for a pure and maximally doped sample ($x=x_0$). $\Delta T_\mathrm{c}$ denotes a change of $\Delta T_\mathrm{c}$ compared to a pure sample. The center horizontal line separates positive (top) and negative (bottom) chemical pressure, respectively. 
    The columns of $N(E_{\mathrm{F}})_{\mathrm{exp}}$ and $N(E_{\mathrm{F}})_{\mathrm{DFT}}$ represent whether the DOS at the Fermi level is determined by experiments or DFT calculations, respectively. Note that 4$H_{\mathrm{b}}$-Ta(S$_{1-x}$Se$_x$)$_2$ exhibits a decrease of $T_{\mathrm{c}}$ above $x_0=30\%$.}
    \label{tab:Tc_doping}
\end{table}

\end{widetext}
\end{document}